\documentclass[prb,twocolumn,superscriptaddress,preprintnumbers,amsmath,amssymb]{revtex4}
\usepackage{graphicx}
\usepackage{amsmath}
\usepackage{amssymb}
\usepackage{epsfig}
\usepackage{wasysym}
\usepackage{bbm}
\usepackage{multirow}
\usepackage{amsmath,amsthm,amssymb}
\renewcommand{\narrowtext}{\begin{multicols}{2} \global\columnwidth20.5pc}

\def\be{\begin{eqnarray}}
\def\ee{\end{eqnarray}}
\newcommand{\nn}{\nonumber\\}

\newcommand{\beq}{\begin{equation}}
\newcommand{\eeq}{\end{equation}}
\newtheorem{thm}{Theorem}[section]
\newtheorem{cor}[thm]{Corollary}

\newtheorem{df}{Definition}[section]

\newcommand{\bpf}{\begin{proof}}
\newcommand{\epf}{\end{proof}}

\begin{document}
\title{The Riemann-Roch Theorem and Zero Energy Solutions of the Dirac Equation on the Riemann Sphere}
\author{Geoffrey Lee }
\address{20 Arlmont Drive, Kensington, CA 94707, USA}

\begin{abstract}
\indent In this paper, we revisit the connection between the
Riemann-Roch theorem and the zero energy solutions of the
two-dimensional Dirac equation in the presence of a delta-function
like magnetic field. Our main result is the resolution of a paradox
- the fact that the Riemann-Roch theorem correctly predicts the
number of zero energy solutions of the Dirac equation despite
counting what seems to be the wrong type of functions.
\end{abstract}

\date{\today}
\maketitle
\tableofcontents
\section{Introduction}
The cross fertilization between mathematics and physics has proven
itself fruitful over the history of science. Often important results
in one field generate new ideas in others. For example, the seminal
Atiyah-Singer index theorem in mathematics has spawned many novel
directions of research in physics due to its connection with the
Dirac equation with a background magnetic field. Examples include:
string theory, particle theory, condensed matter theory, etc. While
solving the Dirac equation with magnetic fields over arbitrary
Riemann surfaces is an arduous task, the Riemann-Roch theorem, which
counts functions of a particular type (namely, meromorphic
functions), correctly predicts the analytic index of the Dirac
Hamiltonian. When the magnetic field is sufficiently strong, this
index agrees with the multiplicity of the zero energy solution.
Therefore it is natural to suspect that the Riemann-Roch theorem can
yield information for the $E=0$ eigenfunctions of the solutions to
the Dirac equation.

\indent The celebrated Riemann-Roch theorem\cite{miranda} deals with
functions that are analytic everywhere except a finite set of poles
on a closed and compact Riemann surface $X$ (a Riemann surface is a
smooth, orientable surface). These functions of interest are known
as meromorphic functions. The characteristics of meromorphic
functions are summarized by a ``divisor''  $D=\sum_{p\in X} n_p~p$,
where $\{p\}$ is a finite (and therefore discrete) set of points on
$X$ and $n_p$ are integers \cite{miranda}. The vector space $L(D)$
consisting of meromorphic functions that have poles of order $\leq
|n_p|$ at the points with $n_p<0$, and zeros of order $\ge n_p$ at
for the points with $n_p\ge 0$ is the entity of interest. The
Riemann-Roch theorem states that
$\dim(L(D))-\dim(L(K-D))=\deg(D)+1-g$ with $K$ any ``canonical
divisor'', $g$ is the genus of $X$ and $\deg(D)=\sum_{p\in X}n_p$,
the degree of the divisor. The concept of the canonical divisor is
irrelevant for the purpose of this paper; this is because for
divisors with sufficiently large degree, $\dim(L(K-D))=0$, hence $K$
does not enter. This is the situation we will be focusing on;
however for further reading about canonical divisors, we refer the
interested reader to Miranda or Hartshorne\cite{miranda}.

\indent In physics it is known that the Riemann-Roch theorem is
connected to the analytical index\cite{atiyah} of the
two-dimensional Dirac equation under a constant magnetic field
\cite{pnueli,iengo}. When the total number of flux quanta\cite{fq}
is sufficiently big, the latter reduces to the degeneracy of the
$E=0$ energy level. In this paper we show that this connection is
quite subtle. First of all, meromorphic functions can diverge while
the solutions\cite{note} of the Dirac equation in uniform magnetic
field do not. This discrepancy can easily be removed by deforming
the magnetic field so that \beq B(\vec{r})=\sum_i 2\pi
n_i\delta(\vec{r}-\vec{r}_i)\label{magnetic field}\eeq while
maintaining the total magnetic flux. After such deformation it is
possible for the $E=0$ solution of the Dirac equation to diverge at
a subset of $\{\vec{r}_i\}$\cite{note2}. The fact that such
deformation preserves the analytic index of the Dirac operator is a
consequence of the Atiyah-Singer index theorem\cite{atiyah}. However
a discrepancy remains even after this deformation, namely, the
solutions of the Dirac equation depends on both $z$ and its
conjugate $\bar{z}$; these functions are not meromorphic. In this
paper we examine and resolve this discrepancy when $X$ is the
Riemann sphere. This case is particularly simple because for any
divisor $D$ with $\deg(D)\geq 0$ on the Riemann sphere,
$\dim(L(K-D))=0$. Moreover, this is the a relevant example to
realistic physical systems. Nevertheless, we believe that the idea
exposed here can be generalized to other Riemann surfaces with
higher genera.

\section{The Riemann-Roch Problem on the Sphere}
For the sake of completeness, in this section the mathematics of the
Riemann-Roch problem will be introduced.
\subsection{Divisors}
\begin{figure}[tbp]
\begin{center}
\includegraphics[angle=0,scale=.3]{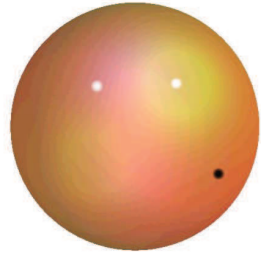}\vspace{-1.3 in}\caption{(color on-line) A graphical representation of a divisor. The white and black points denotes poles and zeros resp.}\label{div}
\end{center}
\end{figure}

In order to assist in the study of meromorphic functions, mathematicians introduce the
concept of divisor. \begin{df} A divisor (Fig. \ref{div}) is a formal finite linear combination of points of $X$ with integral coefficients. The collection of divisors on a surface $X$, denoted $Div(X)$ form an abelian group with the additive group operation defined as though distinct points were linearly independent vectors. Mathematicians define a partial ordering on $Div(X)$ by defining $D\geq 0$ if all its coefficients are nonnegative.\end{df}

\indent Divisors are of importance in the study of Riemann surfaces
because the set of zeros and poles of a meromorphic function on a
surface $X$ can be associated to a divisor.
\begin{df} Let $f$ be a nonzero meromorphic function on a compact Riemann
surface $X$. Then the the divisor of $f$ is defined to be
$div(f)=\sum_{p\in X} n_p p$ where $n_p$ is the order of the zero or
pole of $f$ at $p$. By convention, $n_p$ is positive when $f$ has a
zero at $p$ and negative when $f$ has a pole at $p$. Such divisors
are called principal divisors.\end{df}
It is important to note that such a divisor is always well
defined because a meromorphic function on a compact Riemann surface
can only have finitely many zeros and poles\cite{stein}.

\begin{df} Let $D$ be a divisor over a Riemann surface $X$, we define a complex vector
space $L(D)=\{f {\rm~is~meromorphic~on~}X~|div(f)+D\geq 0\}$
\end{df} The dimension of this space is the quantity of interest.
While in general, it is very difficult to find the functions in
$L(D)$, for the Riemann sphere, we can do so easily.

\subsection{L(D) and its Dimension on the Riemann Sphere}
First, we preform a stereographic projection to map the Riemann
sphere onto $\mathbb{C}_{\infty}:=\mathbb{C}\cup\{\infty\}$, the
one-point compactification of the complex plane \cite{munkres}.
Without loss of generality, we may assume that in any divisor $D$,
the coefficient associated to the point $\infty$ is $0$. This can be
done by appropriately choosing the point from which to
stereographically project.
\begin{figure}[tbp]
\begin{center}
\hspace{-0.8 in.}
\includegraphics[angle=0,scale=.4]{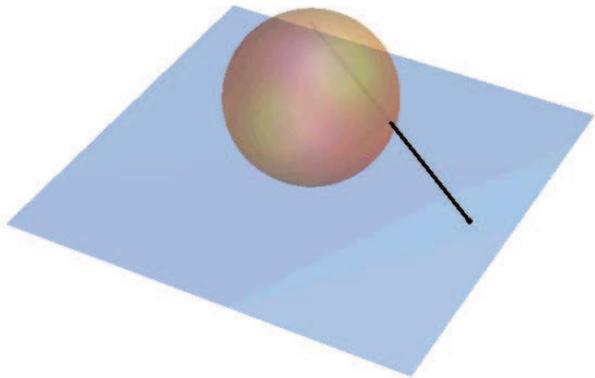}\caption{(color on-line) The stereographic projection. The south pole is
the base point. A point $P$ on the sphere is projected to the plane by finding the intersection of the line connecting the north pole and $P$ and the plane.}\label{steo}
\end{center}
\end{figure}

\begin{thm} Let $D=\sum_{i=1}^N n_ip_i\in
Div({\mathbb{C}_\infty})$ and $\deg(D)\geq 0$. Let
$f_D(z)=\prod_{i=1}^n(z-p_i)^{-n_i}$. Then
$L(D)=\{g(z)f_D(z)|g(z){\rm ~is~a~polynomial~with~} \deg(g)\leq
\deg(D)\}$
\end{thm} For those who are not interested in the proof, read corollary \ref{ld on sphere} before proceeding to the following section.
\bpf We will reproduce the proof in Miranda \cite{miranda} page 149.
Let $g(z)$ be a polynomial with degree $d$. Then $div(g)\geq
-d~\infty$, Hence we have \beq div(f_D)=\sum_i-n_i p_i + (\sum
n_i)\infty .\eeq Therefore \be
&&div(g(z)f_D(z))+D=div(g)+div(f_D)+D\nn&&\geq (\sum_i n_i -d)\infty
= (\deg(D)-d)\infty \ee  which is greater than or equal to $0$ if
$d\leq \deg(D)$. Thus the space $\{g(z)f_D(z)|g(z)$ is a polynomial
with $\deg(g)\leq \deg(D)\}$ is a subspace of $L(D)$. Now, for the
other inclusion, take any nonzero $h\in L(D)$. Then \be
&&div(h/f_D)=div(h)-div(f_D)\nn&&\geq -D-div(f_D)=(-\sum_i
n_i)\infty = -\deg(D)\infty\ee which means that $g$ can only have a
pole of order at most $\deg(D)$ at $\infty$. Thus $g$ is a
polynomial of order at most $\deg(D)$. \epf It is important to note
that in the preceding equations, $\infty$ denotes the antipode of
the point projecting to $0$. This point arises because in general as
$|z|\rightarrow \infty$, the function $f_D(z)$ can have a pole or
zero.
\begin{cor} Let $D\in Div(\mathbb{C}_\infty)$ with $\deg(D)\geq 0$. Then
$\dim(L(D))=1+\deg(D)$. \label{ld on sphere}
\end{cor}
\bpf The set $\{f_D(z)...,z^nf_D(z),...z^{\deg(D)}f_D(z)\}$
generates $L(D)$ as a complex vector space, therefore
$\dim(L(D))=\deg(D)+1$\epf

\section{The Zero Energy Solutions of the Dirac Equation in a Background Magnetic Field}
We will now see the connection between the Riemann-Roch theorem and
the Dirac equation. In order to do so, we first introduce the Dirac
equation on two dimensional surfaces\cite{aharonov}. It is a well
known result from topology that any two dimensional manifold can be
constructed from a planar polygon with edges identified
appropriately\cite{munkres}. Therefore, writing the Dirac equation
over these surfaces amounts to imposing additional periodic boundary
conditions on the corresponding Dirac equation for flat space.

We will use natural units by setting $c=\hbar=1$ as well as fixing
the charge $q$ of the particle so Dirac flux quantum
$2\pi\hbar/qc=2\pi$.

\subsection{The Vector Potential and the Magnetic Field}
Let $B(x,y)$ be a magnetic field. We define the
vector potential $A_x(x,y)$ and $A_y(x,y)$ so that \beq
\partial_x A_y(x,y)-\partial_y A_x(x,y)=B(x,y)\label{gauge}\eeq Eq.(\ref{gauge}) defines the vector potential up to a gauge transformation. The set of vector potentials which
differ from each other by a gauge transformation form an equivalence
class and may be considered the same. Choosing a representative of
such an equivalence class is known as gauge fixing. In the
following, we will fix the gauge by imposing \beq\partial_x
A_y(x,y)+\partial_y A_x(x,y)=0\eeq It will turn out that this choice
of gauge will best exhibit interaction between the Dirac equation
and complex geometry.

\indent With the vector potential defined above, the time
independent Dirac equation in the presence of a magnetic field
becomes\cite{jackiw,aharonov,pnueli}\begin{widetext} \beq \left(
\begin{array}{cc}
 0 & -i\partial_x -A_x-\partial_y +iA_y \\
 -i\partial_x -A_x+\partial_y -iA_y & 0
\end{array}
\right) \left(
\begin{array}{c}
 u(x,y) \\
 v(x,y)
\end{array}
\right) = E \left(
\begin{array}{c}
 u(x,y) \\
 v(x,y)
\end{array}
\right) \eeq\end{widetext}

\subsection{The Magnetic Flux and Divisors}
The type of magnetic field given by Eq.(\ref{magnetic field}) admits a natural representation by a divisor, namely
\beq D=\sum_{i=1}^N n_i p_i \Leftrightarrow
B(\vec{r})=2\pi\sum_{i=1}^N n_i\delta(\vec{r}-\vec{r}_{p_i})\eeq

\subsection{The Vector Potential as a Complex Function}
\begin{figure}[tbp]
\begin{center}
\includegraphics[angle=0,scale=.3]{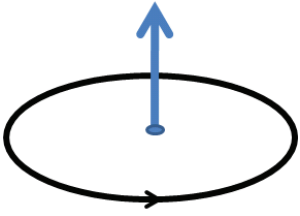}\vspace{-0.5 in}\caption{(color online) As dictated by the Stokes theorem, the loop integral of the vector potential is equal to the magnetic flux. }\label{stokesthm}
\end{center}
\end{figure}

Note that away from the $\vec{r}_i$s, $B(x,y)=0$. Therefore, we have
\begin{eqnarray}&&\partial_xA_y-\partial_yA_x=0\nonumber \\ &&\partial_xA_x+\partial_yA_y=0 \label{two variable CR} \end{eqnarray}
Let us define \beq F(z,\bar{z})=A_y(x,y)+iA_x(x,y)\label{definition of f}\eeq Then Eq.(\ref{two variable CR}) becomes $\partial_{\bar{z}}F(z,\bar{z})=0$, the Cauchy-Riemann condition on $F(z,\bar{z})$. As usual, $\partial_z=(\partial_x-i\partial_y)/2$, and
$\partial_{\bar{z}}=(\partial_x+i\partial_y)/2$. Moreover, by the Stokes theorem, the equation
$\partial_xA_y-\partial_yA_x=2\pi\sum_i
n_i\delta(\vec{r}-\vec{r}_i)$ rewrites into the integral form\beq
\oint_{C_i}\vec{A}(x,y)\cdot d\vec{r}=2\pi n_i\eeq where $C_i$ is a
counter-clockwise loop enclosing only the $i$-th magnetic flux
point (Fig. \ref{stokesthm})

This integral equation translates into \beq \oint_{C_i}F(z)dz=2\pi
in_i \eeq Now, by the Cauchy residue theorem\cite{stein,miranda}, we
can represent $F(z)$ as \beq F(z)=\sum_{i=1}^N \frac{n_i}{z-z_i}\eeq
where $z_i$ represents the position of the $i$-th magnetic flux
point. The reader might have noticed that a holomorphic function
$G(z)$ could be added to $F(z)$ while still preserving the integral.
Ignoring the holomorphic part of $G(z)$ amounts to a further fixing
of gauge.

\subsection{The Dirac Equation in Complex Coordinates}
We first make the following standard definitions:
\begin{eqnarray}
&&A_z=\frac{1}{2}(A_x-iA_y)\nonumber
\\ &&A_{\bar{z}}=\frac{1}{2}(A_x+iA_y)\end{eqnarray} Using $z$ and $\bar{z}$ as
coordinates, the Dirac equation in the presence of a magnetic field
reads \beq 2i\left(
\begin{array}{cc}
 0 & \partial_z-iA_z \\
 \partial_{\bar{z}}-iA_{\bar{z}} & 0
\end{array}
\right)\left(
\begin{array}{c}
 u(z,\bar{z}) \\
 v(z,\bar{z})
\end{array}
\right)=E\left(\begin{array}{c}
 u(z,\bar{z}) \\
 v(z,\bar{z})
\end{array}\right)\label{dirac}\eeq

Taking $E=0$, Eq.(\ref{dirac}) becomes:
\begin{eqnarray} && (\partial_{\bar{z}}-iA_{\bar{z}})u(z,\bar{z})=0
\nonumber \\ && (\partial_z - iA_z)v(z,\bar{z})=0\end{eqnarray}
Substituting in Eq.(\ref{definition of f}), these two equations become \begin{eqnarray} &&
(\partial_{\bar{z}} + \frac{1}{2}\bar{F}(z))u(z,\bar{z}) = 0
\nonumber\\ && (\partial_z - \frac{1}{2}F(z))v(z,\bar{z}) = 0
\end{eqnarray}
\subsection{Solving the Dirac Equation}
The second line of Eq.(\ref{two variable CR}) ensures that it is
possible to pick a $\Phi(x,y)$ so that \beq
A_x=-\partial_y\Phi~~A_y=\partial_x\Phi,\eeq or equivalently,
$A_z=-i\partial_z\Phi/2$ and
$A_{\bar{z}}=i\partial_{\bar{z}}\Phi/2$. Since $A_z=-iF/2$ and
$A_{\bar{z}}=i\bar{F}/2$, this means
\begin{eqnarray}&&{1\over 2}\sum_{i=1}^N \frac{n_i}{\bar{z}-\bar{z}_i}=\frac{1}{2}\bar{F}(z)=\partial_{\bar{z}}\Phi\nonumber\\
&&{1\over 2}\sum_{i=1}^N
\frac{n_i}{z-z_i}=\frac{1}{2}F(z)=\partial_z\Phi \end{eqnarray} In
writing down the above equations we have used Eq.(\ref{definition of
f}).

A simple calculation shows that \beq\Phi(z,\bar{z})=\sum_{i=1}^N
\log\left(|z-z_i|^{n_i}\right)\eeq satisfies the above equations.
Substituting $\Phi$ into the Dirac equation allows us to write the
two differential equations as \begin{eqnarray}
&&(\partial_{\bar{z}}+\partial_{\bar{z}}\Phi(z,\bar{z}))u(z,\bar{z})=0\nonumber
\\
&&(\partial_z-\partial_z\Phi(z,\bar{z}))v(z,\bar{z})=0\label{eq}\end{eqnarray}
Solving this equation by separation of variables gives the solutions
\begin{eqnarray}&&u(z,\bar{z})=f(z)e^{-\Phi(z,\bar{z})} \nonumber \\
&& v(z,\bar{z})=g(\bar{z})e^{\Phi(z,\bar{z})}\end{eqnarray} Notice
that the functions $f(z)$ and $g(\bar{z})$ do not contribute to differentiation with respect to $\bar{z}$ and
$z$ respectively. Substituting in $\Phi(z,\bar{z})$ gives
\begin{eqnarray}&& u(z,\bar{z})=f(z)\prod_{i=1}^N\frac{1}{|z-z_i|^{n_i}}
\nonumber \\ &&
v(z,\bar{z})=g(\bar{z})\prod_{i=1}^N|z-z_i|^{n_i}\label{uv}\end{eqnarray}

Because our original divisor does not specify any poles or zeroes at
$\infty$, $v(z,\bar{z})$ is not a valid solution because it diverges
as $|z|\rightarrow\infty$. Now in order for $u$ to remain bounded at
infinity, $f(z)$ must be a polynomial satisfying $\deg(f(z))\leq
\sum_i n_i$. The reason that $f(z)$ cannot be a rational function of
the form, say $1/z^m$, is because it makes the divisor smaller under
the partial ordering of $Div(X)$ (see ``Divisors''). Thus we
conclude that there are $1+\sum_i n_i$ linearly independent
solutions obtained by setting $f(z)$ equal to $1,z,z^2...z^{1+\sum_i
n_i}$. Notice that $1+\sum_i n_i=\deg(D)+1$, a number which we
calculated to be the complex dimension of $L(D)$ from a purely
mathematical viewpoint in the previous sections.

Despite the above success, the function
$u(z,\bar{z})$ in Eq.(\ref{uv}) is not meromorphic hence is not a member of $L(D)$.
Another equivalent statement is that Eq.(\ref{eq}) is not the Cauchy-Riemann equation
$\partial_{\bar{z}}u=0$.  We will remedy this situation with the use of a
phase transformation.
\section{The phase Transformation}
Our objective now is to transform functions
$u(z,\bar{z})=f(z)\prod_{i=1}^N\frac{1}{|z-z_i|^{n_i}}$ into
meromorphic functions. This can be done by multiplying $u$ by the
unit modulus complex valued function
$\chi(z,\bar{z})=\prod_{i=1}^N(\frac{\bar{z}-\bar{z}_i}{z-z_i})^{n_i/2}$.
By taking the product $\chi u$, we get the meromorphic function \beq
\chi(z,\bar{z})
u(z,\bar{z})=f(z)\prod_{i=1}^N\frac{1}{(z-z_i)^{n_i}}.\eeq Where
$f(z) = 1,z,...,z^{\sum n_i}$. Notice that this gives exactly the
space $L(D)$ of Theorem 2.1.
\subsection{Monodromy}
\begin{figure}[tbp]
\begin{center}
\includegraphics[angle=0,scale=.35]{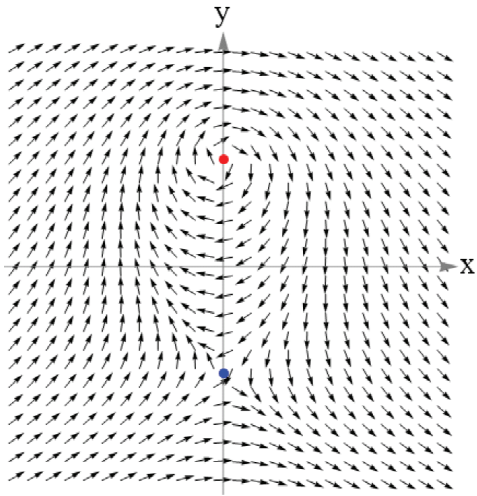}\caption{(color on-line) A plot of the real and imaginary parts of $\chi(z,\bar{z})$ as a vector field.The $n_k$ associated with the red/blue points are $\pm 1$.} \label{monodromy}
\end{center}
\end{figure}

We now show $\chi$ is single valued. It is in fact a nontrivial
result that $\chi$ does not have any branches. Because $\chi$ has
unit modulus, we can write it as $e^{i\theta(z,\bar{z})}$. After
doing so, it suffices to study the behavior of this function along a
loop around any one of the $z_k$. \begin{thm} As $\chi(z,\bar{z})$
wraps once around any one of the $z_k$, $\theta(z,\bar{z})$ changes
by $-2\pi n_k$ (an example is shown in Fig.\ref{monodromy}).
\end{thm}\bpf First it is clear that if the loop encloses only $z_k$, the only factor in $\chi$ that will exhibit
winding is $(\frac{\bar{z}-\bar{z}_k}{z-z_k})^{n_k/2}$. Now, we have
\begin{eqnarray} && \bar{z}-\bar{z}_k=|z-z_k| e^{i\arg(z-z_k)}\nonumber\\
 && z-z_k=|z-z_k| e^{-i\arg(z-z_k)}\end{eqnarray} Therefore, we have
 \beq\left(\frac{\bar{z}-\bar{z}_k}{z-z_k}\right)^{n_k/2}=e^{-in_k\arg(z-z_k)}\eeq Now clearly as $z$
 wraps around $z_k$ once counterclockwise, the exponent of the right hand side changes by $-2\pi n_k$\epf Now we can demonstrate that this phase transformation is single
valued. \begin{cor} $\chi(z,\bar{z})$ has no branches.
\end{cor}\bpf After a full loop $\Delta \theta = -2\pi
n_k$, so $e^{i\Delta\theta}=1$ thus after a loop, the value of
$\chi(z,\bar{z})$ remains the same. \epf

\subsection{Recovery of
Cauchy-Riemann Equations from the Dirac Equation} We will now show
that the transformation of $u(z,\bar{z})$ into a meromorphic
function is not a coincidence. It results from the fact that the
Dirac equation transforms into the Cauchy-Riemann equations after
the phase transformation.
\begin{thm} The equation
$(\partial_{\bar{z}}+\partial_{\bar{z}}\Phi)u=0$ transforms into
$\partial_{\bar{z}}(\chi u)=0$.\end{thm}\bpf Let
$\zeta(z)=\chi(z,\bar{z})u(z,\bar{z})$. Then
$u=\chi^{-1}\zeta=\bar{\chi}\zeta$. We have
\beq\partial_{\bar{z}}u=\partial_{\bar{z}}(\bar{\chi}\zeta)=\zeta\partial_{\bar{z}}\bar{\chi}+\bar{\chi}\partial_{\bar{z}}\zeta\eeq
By an elementary computation
$\chi\partial_{\bar{z}}\bar{\chi}=-\partial_{\bar{z}}\Phi$. As a
result, the Dirac equation becomes \beq
\zeta\partial_{\bar{z}}\bar{\chi}+\bar{\chi}\partial_{\bar{z}}\zeta+(\partial_{\bar{z}}\Phi)(\bar{\chi}\zeta)=0\eeq
Now we can write \beq \zeta\partial_{\bar{z}}\bar{\chi} =
\zeta\bar{\chi} \chi
\partial_{\bar{z}}\bar{\chi} =-\partial_{\bar{z}}\Phi (\zeta\bar{\chi}) \eeq
This cancels the extraneous $\partial_{\bar{z}}\Phi
(\zeta\bar{\chi})$ in the previous equation giving us
the Cauchy-Riemann equation. \epf

\section{Conclusion}

On the Riemann sphere, we have shown that upon a phase
transformation the zero energy solutions of the Dirac equation in
the presence of a magnetic field given by Eq.(\ref{magnetic field})
becomes the meromorphic functions of the Riemann-Roch theorem. We
have also performed the analogous calculation for the complex torus
in which there is a periodic boundary condition requiring us to
represent the vector potential in terms of the elliptic theta
function\cite{miranda}. In this case, the phase transformation is
replaced by the ratio of elliptic theta functions.

In general, for Riemann surfaces of higher genera, it is not clear
that such phase transformations exist. This is because the phase
transformation must be consistent with the appropriate periodic
boundary conditions discussed earlier. A related fact is that it is
difficult to prove the existence of nonconstant meromorphic
functions on arbitrary compact Riemann surfaces; this was first
demonstrated by Riemann\cite{miranda}. We conjecture that the
existence of a phase transformation is equivalent to the existence
of nonconstant meromorphic functions. If this is true, there might
be an alternative proof for the fact that compact Riemann surfaces
are ``projective algebraic''\cite{miranda}.

{\bf Acknowledgements} I thank Prof. Dung-Hai Lee of the University
of California Berkeley for explaining the Dirac equation and its
solutions to me. In addition, I thank Prof. Wu-Yi Hsiang of the
University of California Berkeley for general discussions about
mathematics. I also am grateful to Prof. R. Jackiw for pointing out
relevant references.


\begin{thebibliography}{99}
\bibitem{fq} The number of flux quanta is defined as $\Phi/\Phi_0$ where
$\Phi$ is the total flux passing through the Riemann surface and $\Phi_0$ is the Dirac flux quantum.
We shall choose the natural unit and fix the charge of the particle so that $\Phi_0=2\pi$.
\bibitem{note}In the rest of this paper the phrase "solutions of the Dirac equation" is used to denote the $E=0$ solutions of the time-independent Dirac equation in the presence of magnetic field.
\bibitem{note2} The reader might wonder if deforming
the magnetc flux to make divergent solutions spoils square
integrability. The fact is that the Dirac equation always is a low
energy effective theory, hence has a short-distance cutoff. This
cutoff rounds off the divergence.
\end{thebibliography}
\end{document}